\documentstyle[preprint,aps,tighten,epsf,floats]{revtex}
\newcommand{\ben}{\begin{displaymath}}
\newcommand{\een}{\end{displaymath}}
\newcommand{\be}{\begin{equation}}
\newcommand{\ee}{\end{equation}}
\newcommand{\bea}{\begin{eqnarray}}
\newcommand{\eea}{\end{eqnarray}}

\begin{document}
\draft
\title{Space of State Vectors in ${\cal PT}$ Symmetrical Quantum Mechanics}
\author{ G.S.Japaridze}
\address{Center for Theoretical Studies of Physical Systems, Clark Atlanta
University, Atlanta, GA 30314}
\date{\today}
\maketitle

\begin{abstract}
Space of states of ${\cal PT}$ symmetrical quantum mechanics is examined. Requirement that eigenstates with 
different 
eigenvalues must be orthogonal leads to the conclusion that eigenfunctions belong to the space with an indefinite metric. 
The self consistent expressions for the probability amplitude and average value of operator are suggested. Further specification of space of state vectors yield the superselection rule, redefining notion of the superposition principle.
The expression for the probability current density, satisfying equation of continuity and vanishing for the bound state, is proposed.
\end{abstract}
PACS Numbers: 02.30.Tb, 03.65.Ca,  03.065.Ta
\newpage 

\section{Introduction}
A conjecture of Bessis and Zinn-Justin \cite{bessis} states that the eigenvalues of Schr\"odinger operator with potential $ix^3$ are real and positive. Bender and B\"ottcher \cite{bb} suggested that the 
reason for the absence of complex 
eigenvalues of this non selfadjoint operator could be ${\cal PT}$ symmetry, where ${\cal P}$ is space reflection 
and ${\cal T}$ is time reversal, and using numerical methods and semiclassical approximation, found that spectra of Hamiltonians with potential $ix^N$, $N\geq 2$ are real. The conjecture \cite{bessis}, \cite{bb}, and numerical validation \cite{bb} has provoked a considerable interest in recent years, a sample being refs. \cite{bl}-\cite{zn}. 
Lately, the conjecture of Bessis, Zinn-Justin, Bender and B{\"o}etcher was justified 
using interrelations between the theories of ordinary differential equations and integrable models \cite{patrik}. This approach, based 
on symmetry considerations seems highly promising, paving a road towards identification of non Hermitian Hamiltonians with real spectrum. 

Although the proof of the conjecture that the spectrum of non Hermitian and ${\cal PT}$ invariant
Hamiltonian contains no complex eigenvalues \cite{bb} is still lacking, it is a reasonable question to ask 
whether or not there can exist a self consistent interpretation of the problem described 
by a ${\cal PT}$ invariant Hamiltonian.

Assuming that the spectrum of Hamiltonian under consideration is real, we will pursue the problem of interpretation of
 ${\cal PT}$ symmetrical quantum mechanics. We will take for granted that the "one half of the interpretation", namely that eigenvalues are real is already given, and will concentrate on the "another half" - probabilistic interpretation in terms of the solutions $\psi$.  We will not assume that the solutions of Schr\"odinger equation are eigenfunctions of ${\cal PT}$, i.e. in general  ${\cal PT}\psi(x)\equiv \psi^{\star}(-x)\neq e^{i\omega}\psi(x)$. 

We will consider Schr\"odinger equation in one dimension  with ${\cal PT}$ invariant  
potential  ${\cal PT}V(x)\equiv V^{\star}(-x)=V(x)$ and with non vanishing imaginary part - $ImV(x)\neq 0$. 
We will assume that for Schr\"odinger equation with this $V(x)$ there is no need to invoke analytic continuation in a complex-$x$ plane \cite{bb2}, i.e. the motion is on a real line ${\cal R}:\;-\infty \leq x \leq \infty$. The average values of operators in ${\cal PT}$ invariant field theory 
were investigated in  \cite{bs} using analytic continuation in a complex-$\hat \phi$ plane, where $\hat \phi$ is a field operator. Since we consider the zero dimensional counterpart of nonrelativistic field theory on a real line, the approach based on a Fokker-Planck probability \cite{bs} can not be used in the analysis of the problem under consideration.

The paper is organized as follows. In section $II$, based on the requirement that the state vectors corresponding to the different eigenvalues should be orthogonal,  we will establish that the space of state ${\cal F}$ is the space with an indefinite scalar product
and, as an example of constructing quantum mechanical quantities in  ${\cal PT}$ symmetrical quantum mechanics, we introduce probability current density.

In section $III$ we will overview some necessary material from the theory of indefinite metric spaces, and we will 
find out that the space of states is the special case of spaces with an indefinite metric. Namely, it turns out  
that ${\cal F}$ is the Krein space, decomposable on an orthogonal sums of two Hilbert spaces with positive and negative defined scalar products and allowing to 
introduce a positive defined norm in ${\cal F}$. 

Probability amplitude and average of operators are introduced in section $IV$. It is shown that defining amplitude and average value in terms of vectors belonging to ${\cal F}$ is free from inconsistencies and that the  
Heisenberg operator equations are satisfied. Results are discussed in section $V$.

\section{Orthogonality of State Vectors: Space with an Indefinite Metric}
Combined space reflection and time reversal operator $\theta\equiv  {\cal PT}$ is defined as \cite{landau}:
\begin{equation}
\theta \left\{\ i, \hat x, \hat p \right\}\theta^{-1}=\left\{\ -i,-\hat x,\hat p \right\}
\label{definePT}
\end{equation}
Operator $\hat{A}$ is  $\theta$ invariant if $\theta \hat A \theta^{-1}=\hat A$,
i.e. when $[\hat{\theta},\hat{A}]=0$, the latter valid for the vectors of space in which both $\hat{A}$ and $\hat{\theta}$ can be  simultaneously defined.

We consider Schr\"odinger equation on a real line:
\begin{equation}
\hat {\cal H}\psi(x)=\biggl ( -\frac{\partial^{2}}{\partial x^{2}}+V(x) \biggr )\psi(x)=E\psi(x),
\label{shred}
\end{equation}
with $V^{\star}(-x)=V(x)$, $ImV(x)\neq 0$, and $ImE=0$. Let us address the question about the nature 
of the space of state vectors ${\cal F} \ni \psi$ and about the existence of satisfactory 
interpretation in terms of $\psi$.

As a starting point in analyzing $ {\cal F}$ we consider eigenvalue equations ${\cal H}\psi_{\alpha}=E_{\alpha}\psi_{\alpha}$, ${\cal H}\psi_{\beta}=E_{\beta}\psi_{\beta}$. From
$\theta V(x)\equiv V^{\star}(-x)=V(x)$ it follows that the solutions of (\ref{shred}) are $\psi(x)$ and 
$\psi^{\star}(-x)$ which, in its turn, leads to the relation
\begin{equation}
(E_{\alpha}-E_{\beta})\int_{{\cal R}}dx \psi_{\alpha}(x)\psi^{\star}_{\beta}(-x)=0
\label{shred1}
\end{equation}
In (\ref{shred1})  it is already assumed that $ImE=0$, if eigenvalues are complex, $E_{\alpha}-E_{\beta}$ has to be replaced by $E_{\alpha}-E_{\beta}^{\star}$. 

One of the cornerstones of the interpretation is that
it is impossible to measure two different eigenvalues for the same state  \cite{landau}. Therefore,  
probability is defined in accordance with the requirement that there is no transition 
between the eigenstates with different eigenvalues \cite{landau}.   
In order to maintain in $\theta$ symmetrical quantum mechanics the feature that the transition probability 
between the eigenstates with different eigenvalues vanishes, let us suggest that the transition probability 
amplitude in $\theta$ symmetrical quantum mechanics is 
\begin{equation}
(\psi_{\alpha}|\psi_{\beta})\equiv \int_{{\cal R}}  dx \psi_{\alpha}(x)(\theta \psi_{\beta}(x))=
\int_{{\cal R}}  dx \psi_{\alpha}(x)\psi^{\star}_{\beta}(-x),
\label{inner}
\end{equation}
in other words, we postulate that ${\cal F}$ is a linear space with the scalar product (\ref{inner}). Relations (\ref{shred1})
and (\ref{inner}) imply that $(\psi_{\alpha}|\psi_{\beta})=0$ when $E_{\alpha}\neq E_{\beta}$.

Another way to introduce  the scalar product (\ref{inner}) is as follows. Let both Sturm-Liouville operator $\hat {\cal H}$ and eigenvalue 
$E$  be invariant under the transformation $\Omega$, i.e. let 
$\Omega \hat {\cal H}(x) \Omega^{-1}=\hat {\cal H}(x)$
and $\Omega E=E$. Then, instead  
of starting from (\ref{inner}), one could postulate that the scalar product in ${\cal F}$ 
is defined by:
\begin{equation}
(\psi_{\alpha}|\psi_{\beta})=\int_{{\cal R}} dx \psi_{\alpha}(x)(\Omega \psi_{\beta}(x))
\label{anotherdef}
\end{equation}
When Hamiltonian is Hermitian (when $ImV(x)=0$), definition  (\ref{anotherdef}) leads to the familiar 
expression for the scalar product in a Hilbert space: 
$(\psi_{\alpha}|\psi_{\beta})_{H}=\int_{\cal R} dx \psi_{\alpha}(x)\psi_{\beta}^{\star}(x)\equiv \langle \psi_{\alpha}|\psi_{\beta} \rangle$. The difference between 
$\langle \psi_{\alpha}|\psi_{\beta} \rangle$ and $(\psi_{\alpha}|\psi_{\beta})$ is determined by the symmetry properties 
of Hamiltonian (more precisely, by the symmetry properties of $V(x)$): from the Hermiticity of Hamiltonian it follows that scalar product is 
$\langle \psi_{\alpha}|\psi_{\beta} \rangle$, and for the $\theta$ invariant Hamiltonian  the scalar product, satisfying requirement of orthogonality for $\psi_{\alpha}$ and $\psi_{\beta}$,
is defined as in (\ref{inner}).

When $V(x)$ is $\theta$-invariant and $ImV(x)\neq 0$, $\psi^{\star}(x)$ is not the solution of (\ref{shred}) and as a result  $\langle \psi_{\alpha}|\psi_{\beta} \rangle$ is no longer  orthogonal:
\begin{equation}
(\psi_{\alpha}|\psi_{\beta})_{H}\equiv \langle \psi_{\alpha}|\psi_{\beta} \rangle=\int_{\cal R} dx
\psi(x)_{\alpha}\psi^{\star}_{\beta}(x)\neq 0
\label{notorthog}
\end{equation}
This relation is an evident consequence of $V(x)\neq V^{\star}(x)$.

Since in case of $\theta$ invariant Hamiltonian $\psi_{\alpha}$ and $\psi_{\beta}$ are orthogonal with respect to the scalar product (\ref{inner}), $(\psi_{\alpha}|\psi_{\beta})=0$ when $E_{\alpha}\neq E_{\beta}$,
 one is tempt to interpret the scalar product (\ref{inner}) as the 
transition probability amplitude between the two states described by the
vectors $\psi_{\alpha}$ and $\psi_{\beta}$. This  will lead to a satisfactory result  - transition probability between the states $\psi_{\alpha}$  and $\psi_{\beta}$, labelled by the values of observable $E_{\alpha}\neq E_{\beta}$ is zero, as it  should be for a physical states  \cite{landau}. 

Scalar product (\ref{inner}), respecting orthogonality for the different eigenvalues, is defined in terms 
of $\psi(x)$ and $\psi^{\star}(-x)$. As an example of using $(|)$ instead of the one defined 
in Hilbert space, $\langle|\rangle$,  
let us consider diagonal form $(\psi | V |\psi)$. The remarkable feature is that when  
$V$ is a $\theta$ invariant operator, $(\psi | V |\psi)$ is real:
\begin{eqnarray}
Im(\psi | V |\psi)=Im\int_{{\cal R}}  dx \psi(x)V(x)\psi^{\star}(-x)=\cr\int_{{\cal R}}  dx \Biggl( \biggl[
Re\psi(x)Re\psi(-x)+Im\psi(x)Im\psi(-x)
\biggr]ImV(x)+\cr
\biggl[ Re\psi(-x)Im\psi(x)-Re\psi(x)Im\psi(-x) \biggr]ReV(x) \Biggr) =0,
\label{aver}
\end{eqnarray}
which follows from $ReV(x)=ReV(-x),\;ImV(x)=-ImV(-x)$. Relation (\ref{aver}) resembles the one used 
in quantum mechanics: $Im\langle \psi|V|\psi \rangle=0$ for selfadjoint $V^{\dagger}=V$ \cite{landau}.

As the another example let us examine the following expression:
\begin{equation}
j(x)=\psi(x)\frac{\partial \theta \psi(x)}{\partial x}-\theta \psi(x)\frac{\partial \psi(x)}{\partial x}
\label{current}
\end{equation}
It is straightforward to verify that when $\theta V(x)\equiv V^{\star}(-x)=V(x)$, eq. (\ref{shred}) leads to a 
continuity equation for $j$:
\begin{equation}
\frac{\partial j(x)}{\partial x}=0
\label{conservedcurrent}
\end{equation}
If one uses  $\psi^{\star}(x)$ instead of $\theta \psi(x)$, the continuity equation
 fails: r.h.s. of (\ref{conservedcurrent}) contains $ImV\neq 0$ (since it is assumed that $ImE=0$, we do not consider unstable states).  Symmetry of Hamiltonian dictates firmly that $j(x)$ should be
defined as a bilinear form of $\psi(x)$ and $\theta \psi(x)$. 

Assuming that $j(x)$ represents the probability current density, it is necessary to satisfy besides 
(\ref{conservedcurrent}) another condition, namely that for the bound state $j(x)=0$.
Let us impose on a $\theta$ symmetrical problem (\ref{shred}) the boundary condition, resembling 
the bound state condition for the Hermitian case,
\begin{equation}
\psi(\pm \infty)=0
\label{boundary}
\end{equation}
Note that the well known feature of 
non degeneracy of a one dimensional motion \cite{landau} is still retained - if $\psi_{\alpha}(x)$ and $\psi_{\beta}(x)$ satisfy 
equation (\ref{shred}) and boundary condition (\ref{boundary}) with the same 
eigenvalue, then $\psi_{\alpha}(x)=c\psi_{\beta}(x)$ with $c$ constant
(at this point it is not necessary for eigenvalues to be real). 
When $ImV\neq 0$, the real and imaginary parts of $\psi$ do not satisfy the same equation, so 
 non degeneracy  does not lead to $Im\psi(x)=cRe\psi(x)$. In other words it is not necessary that $\psi=Re\psi+iIm\psi=(1+ic)Re\psi$: solution of the 
Schr\"odinger equation with $\theta$ invariant Hamiltonian, satisfying 
boundary condition (\ref{boundary}), can have a non trivial imaginary part. Since it is $\theta \psi(x)$,
and not $\psi^{\star}(x)$, which satisfies Schr\"odinger equation, non degeneracy implies  that $\psi_{Bound}(x)$, satisfying boundary condition (\ref{boundary}), is the eigenfunction of ${\cal PT}$\footnote{Numerical solution for  $V(x)=ix^3$ shows  that for bound as well as for excited states $Re\psi(x)=Re\psi(-x)$ and $Im\psi(x)=-Im\psi(-x)$, i.e. in this case $\theta\psi(x)=\psi(x)$ \cite{damien}.}:
\begin{equation}
\theta \psi_{Bound}(x)\equiv\psi^{\star}_{Bound}(-x)=e^{i\omega}\psi_{Bound}(x)
\label{boundstate}
\end{equation}
From the definition (\ref{current}) and the relation (\ref{boundstate}) we obtain that for the 
bound state $j(x)$ vanishes:
\begin{equation}
j_{Bound}(x)=0
\label{boundcurrent}
\end{equation}
Equations (\ref{conservedcurrent}) and (\ref{boundcurrent}) indicate that  $j(x)$ could serve as the probability current density: $j(x)$ is conserved, and $j_{Bound}(x)=0$, as one would expect 
for the probability current density \cite{landau}.

So, the scalar product (\ref{inner}) leads to the results similar to the ones of conventional 
quantum mechanics, 
and one could consider (\ref{inner}) as a necessary ingredient for describing and interpreting quantum mechanical problems with ${\cal PT}$ invariant Hamiltonian.

The  subtlety appears when
one address the question of normalizability. Let us examine the diagonal form:
\begin{equation}
(\psi|\psi)=\int_{{\cal R}}  dx \psi(x)\theta \psi(x)=\int_{{\cal R}}  dx \psi(x)\psi^{\star}(-x)\equiv \int_{{\cal R}}  dx \rho(x)
\label{probab}
\end{equation}
If $(\psi_{\alpha}|\psi_{\beta})$ is understood as the transition probability amplitude from the state described by $\psi_{\alpha}$ to the state described by $\psi_{\beta}$, then the
expression  (\ref{probab}) is the amplitude of
probability of the transition from the state characterized by  vector $\psi$  into the same state, 
and the physical requirement is $(\psi|\psi)=1$.

The integrand in (\ref{probab}) has a non zero imaginary part but since $Im\rho(-x)=-Im\rho(x)$ we obtain readily that $(\psi|\psi)$  is real:
\begin{equation}
(\psi|\psi)=\int_{\cal R} dx \Biggl( Re\psi(x)Re\psi(-x)+
Im\psi(x)Im\psi(-x) \Biggr )
\label{probab1}
\end{equation}
The distinctive feature is that the expressions (\ref{probab}), (\ref{probab1}) are not positive defined, and thus $(\psi|\psi)$ can not be normalized to $1$. Positive defined expression is achieved  only when $\psi$ is an even function: $\psi_{ev}(-x)=\psi_{ev}(x)$:
\begin{equation}
(\psi_{ev}|\psi_{ev})=\int_{{\cal R}}  dx \psi_{ev}(x)\psi^{\star}_{ev}(x)\geq 0,
\label{psi+}
\end{equation}
but in general, the diagonal form (\ref{probab}) can be positive, or negative, or zero.
E.g. when $\psi_{\omega}$ is an eigenfunction, i.e. 
$\theta \psi_{\omega}(x)\equiv \psi^{\star}_{\omega}(-x)=e^{i\omega}\psi_{\omega}(x)$, we have:
\begin{equation}
(\psi_{\omega}|\psi_{\omega})=\int_{{\cal R}}  dx \Biggl ( cos\omega \biggl [Re^{2}\psi_{\omega}(x)-Im^{2}\psi_{\omega}(x) \biggr ] -2sin\omega Re\psi_{\omega}(x)Im\psi_{\omega}(x) \Biggr )
\label{badnorm}
\end{equation}
Therefore, ${\cal F}$ is a linear space with indefinite metric,  in particular, 
 from $(\psi|\psi)=0$ it does not necessarily follows that $\psi=0$.

In \cite{bb2} it was suggested that the norm (in a complex $x$-plane) is:
\begin{equation}
\int_{C}dx \psi^{2}(x),
\label{complex}
\end{equation}
and it was conjectured that in the momentum space this norm could be positive defined. Since we are considering motion on a real line ${\cal R}$, using Fourier transform it is straightforward to demonstrate that 
\begin{equation}
(\psi|\psi)=\int_{{\cal R}_{p}} dp \tilde \psi(p)\tilde \psi^{\star}(-p)
\label{momentum}
\end{equation}
i.e. expressions  (\ref{probab}) and (\ref{complex}) ( (\ref{complex}) is the special case of (\ref{probab}), realized when $\theta \psi(x)=\psi(x)$) are not positive defined on the momentum real line as well. 

So, ${\cal F}\in \psi$ is the space with an indefinite metric and we need to specify the space where scalar product is defined via (\ref{inner}) and at the same time it is possible to realize the probabilistic interpretation of $\theta$ symmetrical quantum mechanical problem.
To do so, let us recall some basic statements and theorems from the theory of spaces with an indefinite metric \cite{krein}.

\section{Normalization of State Vectors: Krein Space}
For any element $\psi$ of indefinite metric space ${\cal F}$ there are three possibilities: vector is 
positive $\left\{\psi^+\in {\cal F}^{++}: \;(\psi^+|\psi^+)>0\right\}$, or negative: $\left\{\psi^-\in {\cal F}^{--}:\;(\psi^-|\psi^-)<0\right\}$, or neutral: $\left\{\psi^0\in {\cal F}^0,\;\psi^0\neq 0:\;(\psi^0|\psi^0)=0\right\}$ (clearly the zero vector $\psi=0$ is neutral). In general, ${\cal F}$ contains all three subspaces. 
The semidefinite subspaces ${\cal F}^+$ and ${\cal F}^-$ are defined as the ones with 
nonnegative and nonpositive scalar products: 
 $\left\{\psi^+\in {\cal F}^{+}: \;(\psi^+|\psi^+)\geq 0\right\}$ and 
$\left\{\psi^-\in {\cal F}^{-}: \;(\psi^-|\psi^-)\leq 0\right\}$.

In the semidefinite subspace the scalar product 
$(\psi_{\alpha}|\psi_{\beta})$ is insensitive to $\psi^0$. To see this, let us use the Schwarz inequality, valid in ${\cal F}^{\pm}$  \cite{krein}:
\begin{equation}
|(\psi^0|\psi^{\pm,\;0})|^2\leq (\psi^0|\psi^0)(\psi^{\pm,\;0}|\psi^{\pm,\;0}),
\label{koshi}
\end{equation}
From (\ref{koshi}) it follows that $\psi^0$ is orthogonal to any $\psi\in {\cal F}$: $(\psi^0|\psi^{\pm,\;0})=0$. Therefore, neutral vector does not affect the value of the scalar product: 
$(\psi_{\alpha}+a\psi^0|\psi_{\beta}+b\psi^0)=(\psi_{\alpha}|\psi_{\beta})$. Since $(\psi^0|{\cal F}^{\pm})=0$ and 
$(\psi^0|\psi^0)=0$, below we will consider $\psi^+\in {\cal F}^+$ and $\psi^-\in {\cal F}^-$, pinning out the subspace 
${\cal F}^0$ from the space of states of $\theta$ symmetrical quantum mechanics. Therefore, the first constraint we impose on the space of state vectors is that ${\cal F}$ is an indefinite metric space 
not containing neutral vectors $\psi^0$.

Second constraint originates from the fact that, in general, ${\cal F}$ might be not decomposable as an 
orthogonal sum of ${\cal F}^+$ and ${\cal F}^-$ \cite{krein}, and 
this is the reason why it is impossible to introduce the norm into the whole space ${\cal F}$. 
Space with an indefinite metric ${\cal F}$ can be decomposed as an orthogonal sum of ${\cal F}^+$ and ${\cal F}^-$ when ${\cal F}^+$ and ${\cal F}^-$ are orthogonal regard to the scalar product defined into the whole ${\cal F}$:
\begin{equation}
({\cal F^+}|{\cal F^-})=0
\label{def2}
\end{equation}
In this case, i.e. when  
${\cal F}= {\cal F}^{+}\oplus {\cal F}^{-}$, subspaces  ${\cal F}^{\pm}$ can be completed to Hilbert spaces with the norms 
$\|\psi\|=\sqrt{(\psi|\psi)}$ when $\psi\in {\cal F}^+$, and $\|\psi\|=\sqrt{-(\psi|\psi)}$ when $\psi\in {\cal F}^-$. This is a definition of a Krein space, the indefinite metric space which admits an orthogonal decomposition in which ${\cal F}^{\pm}$ are complete, and  where the positive defined norm can be introduced \cite{krein}.  Based on these 
properties, we suggest that the space of states of $\theta$ symmetrical quantum mechanics is the Krein space.

Let us describe the prescription for introducing norm in the Krein space \cite{krein}.
Define projection operators $\Pi^{\pm}$ satisfying relations:
\begin{equation}
\Pi^{\pm}{\cal F}={\cal F}^{\pm};\;\Pi^++\Pi^-=1;\;\Pi^+\Pi^-=0
\label{def3}
\end{equation}
Operators $\Pi^{\pm}$, $\Pi^{\pm}\psi=\psi^{\pm}$ can not be introduced in any space with an indefinite metric: the necessary condition is the relation (\ref{def2}), i.e. ${\cal F}$ has to be the Krein space (it can be proved that any space with an indefinite metric and positive defined norm can be mapped to a Krein space \cite{krein}).

Next step is to bring into consideration the linear and unitary operator $J$ which maps ${\cal F}$ onto itself: 
${\cal F} \buildrel J \over \longrightarrow {\cal F}$:
\begin{equation}
J\equiv \Pi^+-\Pi^-
\label{def4}
\end{equation}
It can be demonstrated that $J$ is bounded self adjoint operator, and ${\cal F}^{\pm}$ is the eigenspace of $J$ with eigenvalues $\pm 1$. 

Operator $J$ enables to introduce a (positive defined) scalar product $(\psi|\phi)_{{\cal F}}$ into the whole Krein space (i.e. for all 
$\psi,\phi\in {\cal F}$) according to the formula
\begin{eqnarray}
(\psi|\phi)_{{\cal F}}\equiv(J\psi|\phi)=(\psi^+|\phi^+)-
(\psi^-|\phi^-)
\label{aziz}
\end{eqnarray}
Note that when $\psi\in {\cal F}^+$ and $\phi\in {\cal F}^-$, i.e. when $(\psi|\phi)=0$ (see eq. (\ref{def2})),  it follows from (\ref{aziz}) that $(\psi|\phi)_{{\cal F}}=(\psi|0)-(0|\phi)=0$.

Let us now apply prescription (\ref{aziz}) to the indefinite metric space  where the scalar product is given by (\ref{inner}):
\begin{equation}
(\psi|\phi)=\int_{{\cal R}}  dx \psi(x)(\theta \phi(x))=\int_{{\cal R}}  dx \psi(x)\phi^{\star}(-x),
\label{inner1}
\end{equation}
and ${\cal F}={\cal F}^{+}\oplus {\cal F}^{-}$.

To find operator $J$ and thus to define the norm for the problem under consideration, we introduce
\begin{equation}
K^{\pm}\equiv \frac{1\pm {\cal P}}{2},
\label{defk}
\end{equation}
where ${\cal P}$ is the space reflection operator: ${\cal P}\psi(x)=\psi(-x)$. From
 ${\cal P}(K^+\psi(x))=K^+\psi(x)$
we obtain 
\begin{eqnarray}
(K^{+}\psi | K^{+}\psi)=\int_{{\cal R}}  dx K^+\psi(x)(K^+\psi(-x))^{\star}=\cr 
\int_{{\cal R}}  dx K^+\psi(x)({\cal P}(K^+\psi(x)))^{\star}=\int_{{\cal R}}  dx K^+\psi(x)(K^+\psi(x))^{\star}> 0,
\label{decomposition}
\end{eqnarray}
i.e. when the scalar product is given by (\ref{inner}) it follows that the positive vector $\psi^+$ is $\psi^+=K^+\psi$, and the comparison with (\ref{def3}) gives
\begin{equation}
\Pi^+=K^+
\label{k1}
\end{equation}
The arguments similar to those leading to (\ref{k1}), result in $\Pi^-=K^-$, and from (\ref{def4}) and  (\ref{defk}) 
we obtain that in our case $J$ is the operator of space reflection:
\begin{equation}
J=\Pi^+-\Pi^-=K^+-K^-={\cal P}
\label{J}
\end{equation}

Now from  (\ref{aziz}), (\ref{inner1}), and (\ref{J}) it follows that the positive defined scalar product into the whole Krein space ${\cal F}={\cal F}^{+}\oplus {\cal F}^{-}$ is
\begin{equation}
(\psi|\phi)_{{\cal F}}\equiv(J\psi|\phi)=(\psi|{\cal P}\phi)=\int_{{\cal R}}  dx \psi(x)({\cal P}\phi(-x))^{\star}=
\int_{{\cal R}}  dx \psi(x)\phi^{\star}(x)\equiv \langle \psi|\phi \rangle,
\label{scalark}
\end{equation}
i.e. the Hilbert space scalar product is reproduced. The norm in Krein space with the scalar product (\ref{inner}) 
is given by
\begin{equation}
\|\psi\|^{2}=(\psi|J\psi)=(\psi|{\cal P}\psi)=
\int_{{\cal R}}  dx \psi(x)\psi^{\star}(x)\equiv \langle \psi|\psi \rangle \geq 0
\label{definitenorm}
\end{equation}
Expression (\ref{definitenorm}) satisfies requirement of normalizability for the vector $\psi$ characterizing physical 
state - since $\|\psi\|=\sqrt{\|\psi\|^{2}}\geq 0$, one can always renormalize: $\psi\rightarrow \psi^{\prime}$ where 
$\|\psi^{\prime}\|^{2}=1$. Note that "moving backwards", i.e. suggesting that according to (\ref{definitenorm}) the amplitude of transition from $\psi_{\alpha}$ to $\psi_{\beta}$ is given by $\langle \psi_{\alpha}|\psi_{\beta}\rangle = \int dx \psi_{\alpha}\psi^{\star}_{\beta}(x)$ will lead to a wrong result - vectors describing the states with different eigenvalues will no longer be orthogonal (see eq. (\ref{notorthog})).
\section{Probability and average value in $\theta$ symmetrical quantum mechanics}
We will not consider the case when Krein space reduces to ${\cal F}^+$ or to ${\cal F}^-$, i.e. when the space 
of states is a Hilbert space. In other words we assume that $ImV(x)$ can not be removed by a similarity 
transformation. The problem when quantum mechanical system with non Hermitian Hamiltonian can be mapped to the problem defined in Hilbert space with a positive defined scalar product is discussed in \cite{kret}.

We suggest the following expression for the amplitude describing transition from 
the state 
$\psi^j_{\alpha}$ to the state $\psi^{j^{\prime}}_{\beta}$:
\begin{equation}
A^{jj^{\prime}}_{\alpha\beta}=\theta A^{j^{\prime}j}_{\beta\alpha}=
\frac{(\psi^j_{\alpha}|\psi^{j^{\prime}}_{\beta})}{\sqrt{(\psi^j_{\alpha}|\psi^j_{\alpha})}
\sqrt{(\psi^{j^{\prime}}_{\beta}|\psi^{j^{\prime}}_{\beta})}}
\label{amplituda}
\end{equation}
In (\ref{amplituda}) $\alpha,\;\beta$ label eigenstates of Hamiltonian, and $j,j^{\prime}=\pm 1$ are the eigenvalues of the operator $J$ (see (\ref{def4})).

Let us discuss  the amplitude (\ref{amplituda}).

If $j\neq j^{\prime}$, then, according to the definition of Krein space as an orthogonal sum 
(see (\ref{def2})), from 
$(\psi^+|\psi^-)=0$ it follows that $A^{+-}_{\alpha\beta}=0$. Note that if we define $A^{jj^{\prime}}_{\alpha\beta}$ not via the scalar product (\ref{inner}) but by the scalar product in Hilbert space, the "$\pm$ orthogonality" is still valid, since we have 
$({\cal F}^{+}|{\cal F}^-)=0$ as well as $\langle {\cal F}^+|{\cal F}^- \rangle=0$. However, if both 
$\psi_{\alpha}$ and $\psi_{\beta}$ belong only to ${\cal F}^+$, or only to ${\cal F}^-$, 
amplitude written in terms of $\langle \psi^j_{\alpha}|\psi^{j^{\prime}}_{\beta} \rangle$ will no longer 
provide orthogonality, i.e. probability of the transition from the state with eigenvalue $E_{\alpha}$ 
to the state with eigenvalue $E_{\beta}$ will not vanish.
The reason for choosing amplitude as in (\ref{amplituda}) is that $(\psi^j_{\alpha}|\psi^{j^{\prime}}_{\beta})$ guarantees orthogonality 
 when $\psi^j_{\alpha}$ and $\psi^{j^{\prime}}_{\beta}$ belong to the same as well as to the different subspaces of $ {\cal F}$.

The fact that space of state vectors does not contain neutral vector defines a superselection rule
in $\theta$ symmetrical quantum mechanics:
if $\psi^+_{\alpha}$ is a state vector and $\psi^-_{\beta}$ is a state vector, 
then $\phi=c_{\alpha}\psi^+_{\alpha}+c_{\beta}\psi^-_{\beta}$ does 
not corresponds to any realizable dynamical system. Since the equation 
$(\phi|\phi)=c_{\alpha}c^{\star}_{\alpha}(\psi^+_{\alpha}|\psi^+_{\alpha})+
c_{\beta}c^{\star}_{\beta}(\psi^-_{\beta}|\psi^-_{\beta})=0$ can have a non trivial solution, 
$\phi$ can not belong to ${\cal F}$. 
In distinct with Hilbert space, the superposition principle acts separately in 
subspaces $\psi\in {\cal F}^+$ and $\psi\in {\cal F}^-$: linear superposition of 
$\psi^+$ and $\psi^-$ is not the element of ${\cal F}$. This superselection rule resembles 
the one in quantum field theory, where the linear superposition of states with different 
quantum numbers (e.g. $\Psi_{proton}+\Psi_{electron}$), or superposition of  
different representations of Poincare group (e.g $\Psi_{spin\;1}+\Psi_{spin\;1/2}$) 
is not a state vector \cite{bog}. We interpret the
 eigenvalues $\pm 1$ of the operator $J$  as a (conserved) quantum numbers in 
$\theta$ symmetrical quantum mechanics; consequently, one can describe 
$\theta$ symmetrical quantum mechanics as a conventional quantum mechanics in a 
$J$ - invariant space (for a notion of $J$ - invariance see \cite{krein}). 

Using  inequality  (\ref{koshi}) we obtain that the amplitude $A^{jj^{\prime}}_{\alpha\beta}$ satisfies requirements similar to those in quantum mechanics \cite{landau}:
\begin{equation}
|A^{jj^{\prime}}_{\alpha\beta}|\leq 1,\;A^{jj}_{\alpha\alpha}=1
\label{goodamplitude}
\end{equation}
In other words, though the space of states contains negative vectors, definition (\ref{amplituda}) does not faces inconsistencies caused by an indefinite metric.

The another point of interest is the average values of operators.
In analogy with the quantum mechanics we suggest that the average value of the operator $\hat {\cal O}$
in the state described by $\psi^j$ is
\begin{equation}
Av(\hat {\cal O})=\frac{(\psi^j|\hat {\cal O}\psi^j)}{(\psi^j|\psi^j)}
\label{operator}
\end{equation}
Let us consider the time derivative of average value (since $\psi^0\notin {\cal F}$, denominator in 
(\ref{operator}) can always be normalized to $1$, or $-1$; derivation below is valid for any of the signs):
\begin{equation}
\frac{d}{dt}Av(\hat {\cal O})=\int_{\cal R}dx\Biggl ( \frac{\partial (\theta \psi(x,t))}{\partial t}\hat {\cal O}\psi(x,t)+
(\theta \psi(x,t))\hat {\cal O}\frac{\partial \psi(x,t)}{\partial t}  \Biggr )
\label{time}
\end{equation}
Using the time dependent Schr\"odinger equation with the $\theta$ invariant Hamiltonian
\begin{equation}
i\frac{\partial \psi^j(x,t)}{\partial t}=\hat {\cal H}\psi^j(x,t),\;-i\frac{\partial (\theta \psi^j(x,t))}{\partial t}=
 \hat {\cal H}(\theta \psi^j(x,t))
\label{timeshred}
\end{equation}
it is straightforward to demonstrate that
\begin{equation}
\frac{d}{dt}Av(\hat {\cal O})=iAv(\hat {\cal H}\hat {\cal O}-\hat {\cal O}\hat {\cal H})
\label{erenfest}
\end{equation}
Therefore in a $\theta$ symmetrical mechanics Heisenberg equation
\begin{equation}
i\frac{d\hat {\cal O}}{dt}=\hat {\cal O}\hat {\cal H}-\hat {\cal H}\hat {\cal O}
\label{heisen}
\end{equation}
is satisfied. Of course, operator equation (\ref{heisen}) is defined in space ${\cal F}$ with the scalar product (\ref{inner}). 

Note that the scalar product (\ref{inner}) satisfies relation 
$(\psi|\hat{\cal O} \phi)=(\theta \hat{\cal O} \theta^{-1} \psi|\phi)$, 
the analogy of which in Hilbert space is 
$\langle \psi|\hat{\cal O} \phi\rangle=\langle \hat{\cal O}^{\dagger}\psi|\phi\rangle$. Since selfadjoint operator in Hilbert space can be not selfadjoint in Krein space, expression (\ref{operator}) can be associated with the values of 
observables  only when $\hat {\cal O}$ is $\theta$ invariant, like Hamiltonian, momentum, or operator $i\hat x$. 
In connection with the notion of self adjoint operators in Krein space let us refer to the following theorem: the 
spectrum of the operator $\hat {\cal O}$ which is symmetric (i.e. $(\hat {\cal O}\psi|\phi)=(\psi|\hat {\cal O}\phi)$ for every 
$\psi,\;\phi\in {\cal F}$) and positive (i.e. $(\hat {\cal O}\psi|\psi)\geq 0$ for every $\psi\in {\cal F}$) is real \cite{krein}. According to 
this theorem, the necessary condition for the operator $\hat{ \cal O}$ to have a real spectrum is that 
$\hat{ \cal O}$ is a symmetric operator, in other words the necessary condition is that operator is 
$\theta$ invariant,  the case we are concerned with. The spectrum will be real if $\theta$ invariant operator 
$\hat {\cal O}$ is positive, the 
case necessary to be investigated separately for every given operator. This problem lies beyond the scope of the 
present paper.
\section{discussion}
We have considered quantum mechanical problem with $\theta\equiv {\cal P}{\cal T}$ invariant Hamiltonian 
$\hat {\cal H}=-\partial^2+V(x)$ with $\theta V(x)\equiv V^{\star}(-x)=V(x),\;ImV(x)\neq 0$. Requiring orthogonality and 
using the symmetry of Hamiltonian we have obtained that the scalar product in space of state vectors ${\cal F}$ is 
given by (\ref{inner}), which, in its turn leads to the conclusion that ${\cal F}$ is a space with an indefinite metric. The requirement of normalizability leads to the further constraints on the space of state vectors and as the result, 
${\cal F}$ can be identified with the Krein space. The latter can be (loosely) defined as the orthogonal 
sum of two Hilbert 
spaces, rigged with positive and negative defined scalar products. Excluding neutral vector 
$\psi^0\neq 0:\;(\psi^0|\psi^0)=0$ from physical states we have arrived to the superselection rule in ${\cal F}$: 
not any superposition of state vectors belongs to ${\cal F}$. Transition amplitude (\ref{amplituda}), and the 
average value (\ref{operator}) can be defined in a self-consistent way in ${\cal F}$.

The superselection rule is the feature characterizing the quantum field theory, where, in distinct with quantum mechanics, there is no one to one correspondence between the pure states and the rays of space where the algebra 
of field operators is realized \cite{bog}. In other words, space of states in quantum field theory is not the linear space, but rather the orthogonal sum of linear spaces \cite{bog}: the same happens to be  true in $\theta$ symmetrical quantum mechanics. 

Another aspect of $\theta$ symmetrical quantum mechanics, not realized 
in conventional quantum mechanics, is the space with an indefinite metric. Again, this feature shows up in a certain 
field-theoretical models: the two well established examples are quantum electrodynamics \cite{bog}, and the exactly solvable field theoretical model by T.D.Lee \cite{lee}. The indefinite metric in quantum field theory is in a sense 
"fictitious", since it corresponds to the dynamics of the redundant degrees of freedom, and the requirement that 
dynamics should be realizable in the space of physical degrees of freedom leads to the reduction of the space of all 
(physical and non physical) states to the space of physical states with positive defined scalar product; the 
space of physical states is 
 complete relative to this scalar product \cite{bog}, \cite{lee}. In a $\theta$ symmetrical quantum mechanics there 
are no "extra, non physical" degrees of freedom, therefore it is impossible to introduce auxiliary condition 
allowing to eliminate
subspace with the non positive defined scalar product. In general, the complete set of base vectors 
in spaces with an indefinite metric consists of vectors with positive, as well as with non positive norm: 
${\cal F}$ is complete, i.e. it is a Hilbert space relative to the norm $\langle\psi|\psi\rangle$ \cite{krein}. 
The reduction 
of space of states can result in to an  incomplete set (the problem of completeness for the potentials $ix^3$ and 
$-x^4$ is discussed in \cite{bb3}) which would make interpretation impossible.

Concluding, it is possible to give a self consistent interpretation for ${\cal PT}$ symmetrical version 
of quantum mechanics. The price one has to pay is to abandon the Hilbert space and to switch to Krein space 
with an indefinite metric. This feature, as well as the superselection rule are the aspects not present in 
formulation of the Hermitian quantum mechanics. Nevertheless, they do not violate the general requirements 
presented to probabilistic interpretation. The results presented are valid when $ImV(x)$ is not vanishing, therefore 
there can not be a smooth transition to the Hermitian case: from the very beginning, depending on the symmetry 
of Hamiltonian, the scalar product is defined either as $(|)$ (Krein space), or as $\langle | \rangle$ 
(Hilbert space).

Needless to say that any discovery of a dynamical system, described in terms of non Hermitian and 
${\cal PT}$ invariant Hamiltonian will be more than welcome.
\begin{flushleft}
{\bf Acknowledgments}
\end{flushleft}
Helpful discussions with C.M.Bender, D.Bessis, C.R.Handy, R.Kretschmer, G.A.Mezincescu,  S.Naboko, L.Szymanowski, and X.Q. Wang  are highly  appreciated. 

While this manuscript was under revision, authors of ref. \cite{Znoj} arrived to the same probability current, as the expression (\ref{current}) of the present paper.

\end{document}